\documentclass[aps,twocolumn,english,superscriptaddress,citeautoscript,showkeys,preprintnumbers,amsmath,amssymb,floatfix,footinbib]{revtex4}
\usepackage{hyperref}
\hypersetup{colorlinks=true,linkcolor=red,citecolor=blue}
\usepackage{amsmath}
\usepackage{graphicx}
\usepackage{dcolumn}
\usepackage{float}

\begin{document}
\title{ Non-Fermi-liquid behavior at anti-ferromagnetic quantum critical point in heavy fermion system Ce(Cu$_{1-x}$Co$_x$)$_2$Ge$_2$}
\author{Rajesh Tripathi}
\affiliation{Department of Physics, Indian Institute of Technology, Kanpur 208016, India}
\author{Debarchan Das}
\affiliation{Department of Physics, Indian Institute of Technology, Kanpur 208016, India}
\affiliation{Max-Planck Institute for Chemical Physics of Solids, 01187 Dresden, Germany}
\author{C. Geibel}
\affiliation{Max-Planck Institute for Chemical Physics of Solids, 01187 Dresden, Germany}
\author{S. K. Dhar}
\affiliation{DCMP and MS, Tata Institute of Fundamental research, Mumbai 400005, India}
\author{Z. Hossain}
\email{zakir@iitk.ac.in}
\affiliation{Department of Physics, Indian Institute of Technology, Kanpur 208016, India}
%\affiliation{Max-Planck Institute for Chemical Physics of Solids, 01187 Dresden, Germany}

\date{\today}

\begin{abstract}
Polycrystalline samples of Ce(Cu$_{1-x}$Co$_x$)$_2$Ge$_2$  were investigated by means of electrical resistivity $\rho$($T$), magnetic susceptibility $\chi$($T$), specific heat $C$$_p$($T$) and thermo electric power $S$($T$) measurements. The long-range antiferromagnetic (AFM) order, which set in at $T$$_N$ = 4.1 K in CeCu$_2$Ge$_2$, is suppressed by non-iso-electronic cobalt (Co) doping at a critical value of the concentration $x$$_c$ = 0.6, accompanied by non-Fermi liquid (NFL) behavior inferred from the power law dependence of heat capacity and susceptibility i.e. $C$($T$)/$T$ and $\chi$($T$) $\propto$ $T$$^{-1+\lambda}$ down to 0.4 K, along with a clear deviation from $T$$^2$ behavior of the electrical resistivity. However, we have not seen any superconducting phase in the quantum critical regime down to 0.4 K.


\end{abstract}

\maketitle

\section{INTRODUCTION}

In some compounds of Ce and Yb, a second order quantum phase transition (QPT) at $T$$\rightarrow$0 separates the ordered and paramagnetic states and leads to interesting properties such as non-Fermi liquid, heavy fermion (HF) behavior and/or unconventional superconductivity. For example, the HF metal CeCu$_2$Si$_2$ and it's sister analogue CeCu$_2$Ge$_2$ both show superconductivity around their AFM quantum critical point (QCP) under pressure \cite{Behnia, DJaccard, Grosche}. At ambient pressure, CeCu$_2$Ge$_2$  is antiferromagnetically ordered heavy fermion system (HFS) with N$\acute{e}$el temperature $T$$_N$ = 4.1 K and a characteristic Kondo lattice temperature $T$$^*$ = 6 K \cite{Boer}, with similar energy scales of Kondo and RKKY interaction. With increasing pressure the hybridization between $4f$ and conduction electrons due to Kondo effect increases, which suppresses antiferromagnetism and eventually superconductivity emerges. The superconductivity around AFM QCP is believed to be mediated by magnetic fluctuations, as inferred from neutron scattering experiments \cite{Knopp}. Superconductivity has also been observed in Ge substituted CeCu$_2$(Si$_{1-x}$Ge$_x$)$_2$ \cite{Knebel} and Ni substituted  Ce(Cu$_{1-x}$Ni$_x$)$_2$Si$_2$ \cite{Yoichi} around the AFM QCP. The quantum critical phenomenon and the associated NFL behavior in such cases arises due to the fluctuations of the AFM order parameter with diverging intensity at the QCP, as described in the spin fluctuation theories of Hertz, Millis and Moriya \cite{Mili}. Although numerous investigations on CeCu$_2$Ge$_2$ have been carried out using high pressure, low temperature and magnetic field, the effect of disorder on the physical properties close to magnetic-nonmagnetic boundary has not been addressed.

The competition between Ruderman-Kittel-Kasuya-Yosida (RKKY) and Kondo interaction in HFS offers the opportunity to tune these systems towards magnetic-nonmagnetic boundary by alloying or hydrostatic pressure. It has been observed that the NFL behavior of some chemically substituted f-electron systems is better described within the context of Castro Neto theory based on Griffiths' singularities \cite{Andrade,Lai,Ghosh,Krishnamurthy,MTran,Lohneysen}. At the QCP, NFL behavior in such systems is phenomenologically found to be described with $C$($T$)/$T$ and $\chi$($T$) $\propto$ $T$$^{-1+\lambda}$, where $\lambda$ is slightly smaller than 1.0 and a power law in the resistivity  $\rho$($T$) = $\rho$$_0$ + $A$$T$$^\alpha$ with either $\alpha$ $\approx$ 1 or 1.5 for 2D and 3D quantum fluctuations respectively \cite{Castro,Andrade,Lohneysen,Moriya}. So far, many HFS belonging to this category (alloying) have been investigated successfully with vanishing AFM phase transitions near QCP e.g. CeCu$_{6-x}$Ag$_x$ \cite{Scheidt}, YbCu$_{5-x}$Al$_x$ \cite{Bauer}, Ti$_{1-x}$Sc$_x$Au \cite{Svanidze}. In Ce(Cu$_{1-x}$Ni$_x$)$_2$Ge$_2$, the $x$-$T$ phase diagram shows a transition from a local moment type of AFM ordering for $x$ $<$ 0.2 to a heavy-fermion band magnetism between 0.2 $\leq$ $x$ $\leq$ 0.75 and finally to a Fermi liquid close to $x$ = 1 \cite{Loidl,Sparn}. Compared to Cu($3d^{10}$), Ni ($3d^9$) has one less electron where as Co($3d^8$) has two less electrons.  Thus, it is expected that Co doping introduces more electronic disorder in the Cu-Ge layer. A preliminary reports on Ce(Cu$_{1-x}$Co$_x$)$_2$Ge$_2$ \cite{Maeda} based only on resistivity and specific heat measurements exists in the literature indicating a possible critical concentration of $x$ = 0.5 - 0.6 for suppression of magnetic order. Here, CeCo$_2$Ge$_2$ is an intermediate valence/heavy fermion compound with relatively high Kondo temperature($T$$_K$) \cite{Fujii}. In the present work, we have carried out a comprehensive study of the low temperature properties of Ce(Cu$_{1-x}$Co$_x$)$_2$Ge$_2$  by means of electrical resistivity $\rho$($T$), magnetic susceptibility $\chi$($T$), heat capacity C$_p$($T$) and thermoelectric power $S$($T$) measurements. Besides making more compositions with various values of $x$ than in ref. \cite{Maeda}, we report the magnetic susceptibility and the thermopower data in this system for the first time. Our results show that the AFM ground state of CeCu$_2$Ge$_2$ can be continuously suppressed by Co doping and around the critical concentration $x$$_c$ $\sim$ 0.6 there are indications of a breakdown of FL behavior, in particular, the heat capacity divided by temperature $C/T$ and $\chi$($T$) diverges with decreasing temperature.

\section{METHODS}

Polycrystalline samples of Ce(Cu$_{1-x}$Co$_x$)$_2$Ge$_2$ for 0 $\leq$ $x$ $\leq$ 1 were prepared by arc melting the constituent elements, taken in proper ratio, in an argon atmosphere. Some of the samples were subjected to heat treatment in evacuated sealed quartz tubes at 850$^\circ$C for one week. We found that the residual resistivity of the homogenized ingots is significantly lower than that of the as-cast specimens. The results presented here were obtained on the annealed  specimens. Powder x-ray diffraction with Cu-K$_{\alpha}$ radiation was used to determine the phase purity and crystal structure. Scanning electron microscope (SEM) equipped with energy dispersive x-ray (EDX) analysis was used to check the homogeneity and composition of the samples. The magnetic measurements in the temperature range 2 - 300 K were carried out using a commercial Vibrating Sample Magnetometer (VSM) attached with physical property measurement system (PPMS, Quantum Design) whereas measurements in the temperature range 0.4 K - 2 K were accomplished in Quantum Design SQUID magnetometer equipped with a Helium-3 option. The specific heat was measured by relaxation method in PPMS. Electrical resistivity measurements  in the temperature range 2 - 300 K  were performed using standard dc transport option of the PPMS.  In addition, electrical resistivities of few selected samples were measured down to 0.35 K by using ac transport option of PPMS. Thermoelectric power (TEP) was measured using thermal transport option (TTO) of PPMS using thermal relaxation method. A heat pulse of 30 seconds was applied to raise the temperature of the hot end by 3\% of the base temperature.

\section{RESULTS AND DISCUSSION}

%\subsection*{\label{ExpDetails} A. Crystal Structure}

\begin{figure}
\includegraphics[width=8.5cm, keepaspectratio]{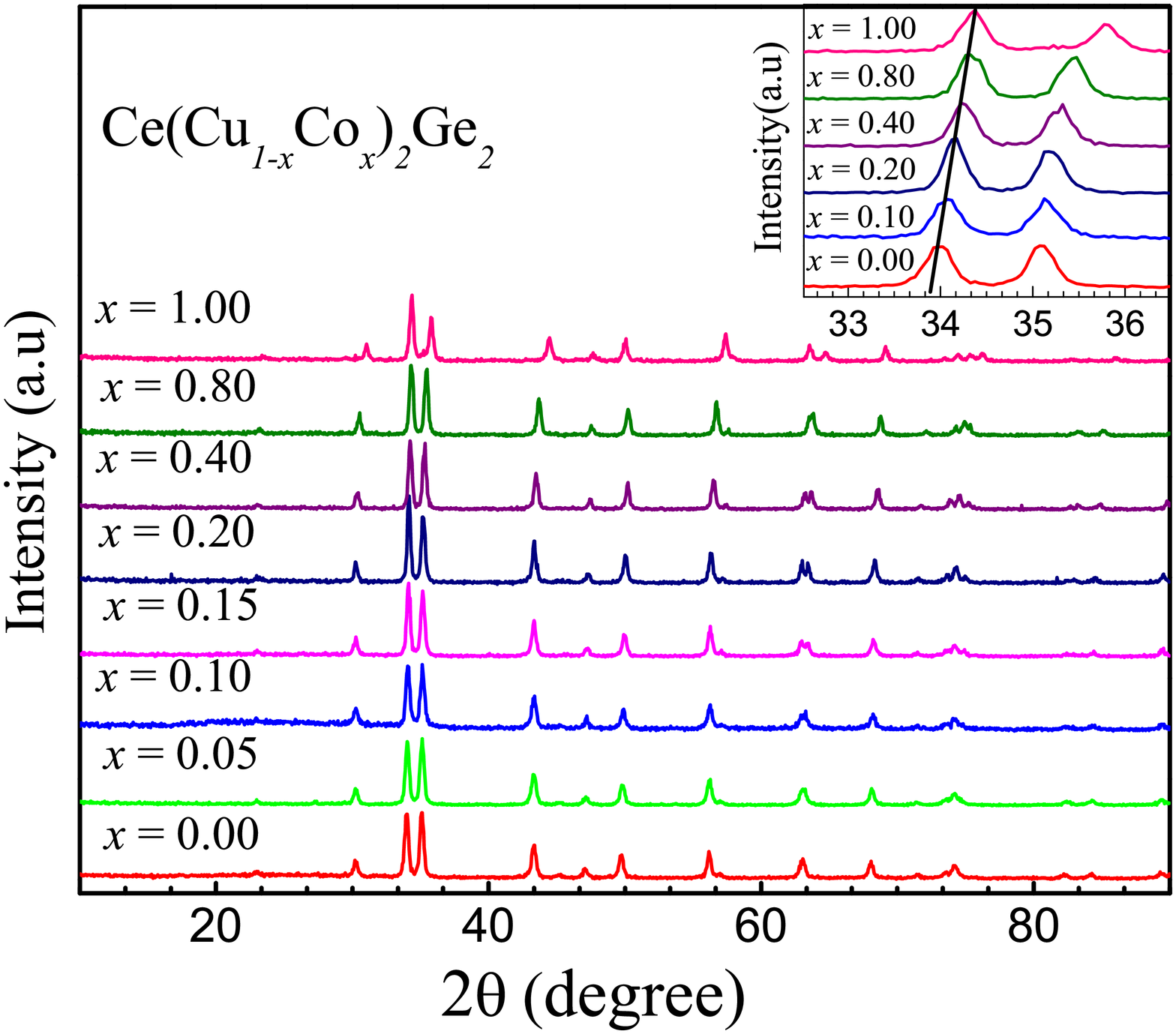}
\caption{\label{fig:CeCuCoGe-XRD} (Color online) Room temperature x-ray diffraction pattern of Ce(Cu$_{1-x}$Co$_x$)$_2$Ge$_2$.  Inset shows the shifting of peaks with Co doping.}
\end{figure}

\begin{figure}
\includegraphics[width=8.5cm, keepaspectratio]{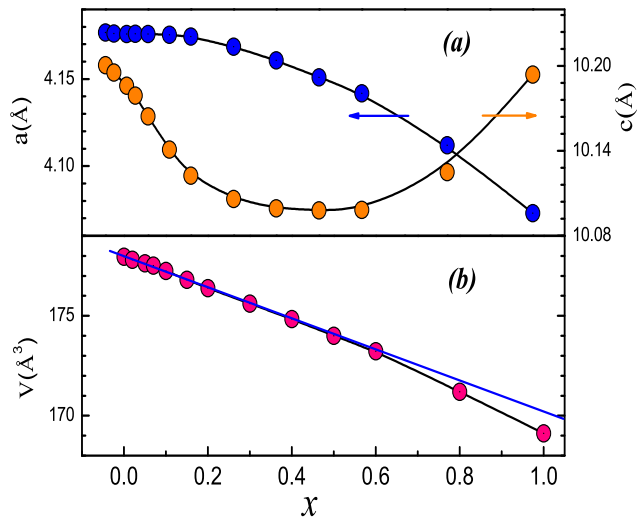}
\caption{\label{fig:CeCuCoGe-LP}(Color online) Lattice parameters(upper panel) and volume(lower panel) of Ce(Cu$_{1-x}$Co$_x$)$_2$Ge$_2$ at room temperature as a function of Co concentration $x$. Solid lines passing through the symbols are guide to the eyes. A change of slope in $V(x)$ is seen around $x$ = 0.6 in lower panel.}
\end{figure}

Powder x-ray diffraction patterns of Ce(Cu$_{1-x}$Co$_x$)$_2$Ge$_2$ for 0 $\leq$ $x$ $\leq$ 1 (Fig.1) confirm that each member of the series is single phase crystallizing in the ThCr$_2$Si$_2$-type tetragonal structure with space group $I4/mmm$. The lattice parameters for $x$ = 0 and $x$ = 1 are in good agreement with the values reported in literature for CeCu$_2$Ge$_2$ \cite{Boer} and CeCo$_2$Ge$_2$ \cite{Takashi} respectively. The lattice volume (Fig.2(b)) is found to decrease continuously for entire $x$  without any change in crystal structure, though the $c$-axis expands beyond $x$ $\sim$ 0.5 (Fig.2(a)). A clear change of slope in the $x$ dependence of lattice volume around $x$ = 0.6 is observed, signalling a change in the cerium valence. The relative change in the volume from $x$ = 0 to 0.6 is about - 2.5\%. The volume contraction results in a chemical pressure which can be calculated using the Birch-Murnaghan equation $P$  = $B_0$$\Delta$$V(x)$/$V(0)$, where $B_0$ is the bulk modulus and its value for CeCu$_2$Ge$_2$ is reported to be 98 GPa \cite{Reul}. The estimated value of chemical pressure thus comes out to be  $P$ = 2.6 GPa for $x$ = 0.6 and $P$ = 4.9 GPa at $x$ = 1.0.

%\begin{table}
%\centering
%\caption {Lattice parameters $a$({\AA}), $c$({\AA}), $c/a$ ratio and unit-cell volume $V$({\AA$^3$}) of ThCr$_{2}$Si$_{2}$-type tetragonal system Ce(Cu$_{1-x}$Co$_x$)$_2$Ge$_2$  ($x$ = 0, 0.02, 0.05, 0.1, 0.15, 0.2, 0.3, 0.4, 0.5, 0.6, and 0.8)}
%\label{table:CeCuCoGe-tableI}
%\vskip .5cm
%\addtolength{\tabcolsep}{+5pt}
%\begin{tabular}{c c c  c c}
%\hline
%\hline
%$x$ & $a$ ({\AA})  &$c$ ({\AA})	& $c/a$			& $V$ ({\AA$^3$})\\[0.5ex]

%\hline
%0 & 4.176(3)	&10.202(1)	& 2.442			    & 177.9\\[1ex]

%0.02 & 4.176(2)	&10.19(1)	& 2.441			    & 177.8\\[1ex]

%0.05 & 4.175(1)	&10.186(7)	& 2.439		        & 177.6\\[1ex]

%0.07 & 4.176(1)	&10.179(8)	&  2.437			& 177.5\\[1ex]

%0.1 &4.175(9)	&10.16(2)	&  2.434			& 177.2\\[1ex]

%0.15 & 4.175(1)	&10.141(7)	&  2.429			& 176.8 \\[1ex]

%0.2 & 4.174(3)	&10.122(1)	& 2.425			    & 176.4\\[1ex]

%0.3 & 4.168(3)	&10.106(1)	& 2.425			    & 175.6\\[1ex]

%0.4 & 4.160(2)	&10.099(1)	&  2.427			& 174.8 \\[1ex]

%0.5 & 4.151(3)	&10.098(1)	& 2.425			    & 173.9\\[1ex]

%0.6 & 4.141(3)	&10.098(1)	& 2.425			    & 173.2\\[1ex]

%0.8 & 4.112(5)	&10.125(1)	&  2.462			& 171.2 \\[1ex]

%1 & 4.073(8)	&10.194(1)	&  2.502			& 169.2 \\[1ex]

%\hline

%\end{tabular}
%\end{table}

%\subsection*{\label{ExpDetails} B. Magnetic Susceptibility}

Magnetization $M$($T$) measurements were carried out for Ce(Cu$_{1-x}$Co$_x$)$_2$Ge$_2$ at a fixed applied field of $H$ = 0.1 $T$ and the resulting susceptibilities $\chi$($T$) = $M$($T$)/$H$ for $x$ = 0, 0.02, 0.05 are plotted in Fig.3. Upper right inset of Fig.3 shows $\chi$($T$) of $x$ = 0.2 and 0.4 down to 0.4 K where as the lower left inset shows inverse susceptibility as a function of temperature for $x$ = 0.4, 0.6, 0.8, and 1. Antiferromagnetic transition temperature also referred as N$\acute{e}$el temperature $T$$_N$ is defined by the pronounced  maxima (indicated by arrows) in $\chi$($T$). $T$$_N$ is found to shift towards low temperature with increasing $x$. For $x$ $\geq$ 0.4 no anomaly due to magnetic ordering is found down to 0.4 K. It is important to note that unlike Ce(Cu$_{1-x}$Ni$_x$)$_2$Ge$_2$ \cite{Sparn,Loidl,ButtgenB} we have not observed a further increase of $T$$_N$ or even two different $T$$_N$ simultaneously for intermediate concentrations down to 0.4 K. At high temperature ($T$ $>$ 200 K), the susceptibility follows modified Curie-Weiss behavior [$\chi$ = $\chi$$_0$ + C/($T$ - $\theta$$_P$)]. Here $\chi$$_0$ is the temperature independent term and C = N$\mu$$_{eff}$$^2$/3$k$$_B$, where $\mu$$_{eff}$ is the effective moment. The Curie-Weiss temperatures $\theta$$_P$ obtained from the fits of the high-temperature (200 K $\leq$ $T$ $\leq$ 300 K) susceptibilities with the above equation for 0 $\leq$ $x$ $\leq$ 1 are presented in table I.  With increasing Co concentration, $\theta$$_P$ increases to a value of -105 K at $x$ = 0.6 and then to even larger negative values of -399 K for $x$ = 1. This is a common feature in Ce-based materials with strong hybridization between the $4f$ and conduction electrons and indicates that the Kondo interaction strengthens with increasing $x$ \cite{Sereni}. Gr\"{u}ner and Zawadowski \cite{Gruner} have shown that the absolute value of $\theta$$_P$ is related to the Kondo temperature as $T$$_K$ = $\mid$$\theta$$_P$$\mid$/4. From this relation we estimated the value of $T$$_K$ (for CeCu$_2$Ge$_2$, $T$$_K$ = 6 K and for CeCo$_2$Ge$_2$, $T$$_K$ = 100 K) which are very similar to those reported in literature \cite{Knopp,Fujii,Venturini}. $T$$_K$ for all concentrations are given in table I. The effective moment of CeCu$_2$Ge$_2$ is found to be 2.50 $\mu$$_{B}$. Furthermore, the effective moment ($\mu$$_{eff}$) for CeCo$_2$Ge$_2$ and some intermediate concentrations are slightly higher than the theoretical value of Ce$^{3+}$ (2.54 $\mu$$_B$ corresponding to the $J$ = 5/2 multiplet of the free Ce$^{3+}$ ion). Therefore, at high  temperature the valance state of Ce is close to Ce$^{3+}$ even for higher $x$ values which is consistent with soft x-ray resonant photoemission investigation \cite{Venturini} and near-edge x-ray absorption study \cite{Ansari} on CeCo$_2$Ge$_2$. Figure 4 shows the temperature dependence of the magnetic susceptibility of Ce(Cu$_{1-x}$Co$_x$)$_2$Ge$_2$ for $x$ = 0.4, 0.6 (where the magnetic order is completely suppressed($T$$_N$$\rightarrow$0)), and 0.8 on logarithmic (both axes) plot. The solid lines in Fig.4 represent the least squares fits of the Castro Neto model i.e. $\chi$($T$) = $\chi$($0$)$T$$^{-1+\lambda_\chi}$, at low temperatures, where $\lambda_\chi$ is a parameter determined by the best fit. The values of $\lambda_\chi$ for different compositions $x$ are given in table II. It is to be noted that the NFL like power law dependence is seen even for $x$ = 0.8 sample. While these results are suggestive of quantum Griffith singularities, further measurements at low temperature are required to verify our conjecture.

\begin{figure}
\includegraphics[width=8.5cm, keepaspectratio]{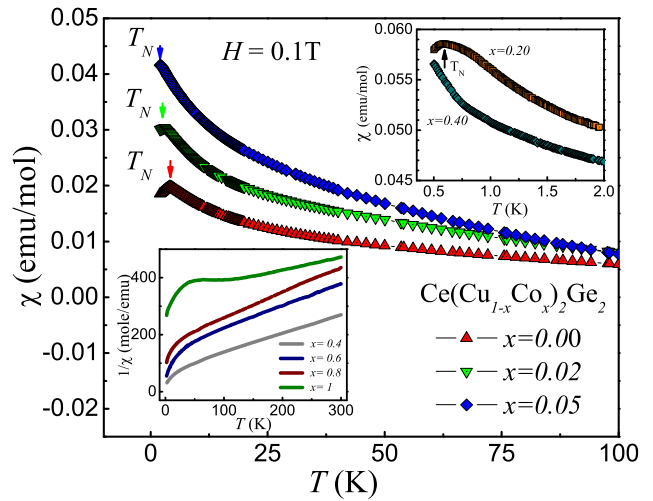}
\caption{\label{fig:CeCuCoGe-SUSP} (Color online) Temperature dependence of the magnetic susceptibility for $x$ = 0, 0.02, and 0.05. The AFM transition temperatures $T$$_N$ are marked by arrows. The upper right inset shows the data for $x$ = 0.2 and 0.4 in the milikelvin temperature range where as the lower left inset shows inverse susceptibility data for $x$ = 0.4, 0.6, 0.8, and 1.}
\end{figure}

\begin{figure}
\includegraphics[width=8.5cm, keepaspectratio]{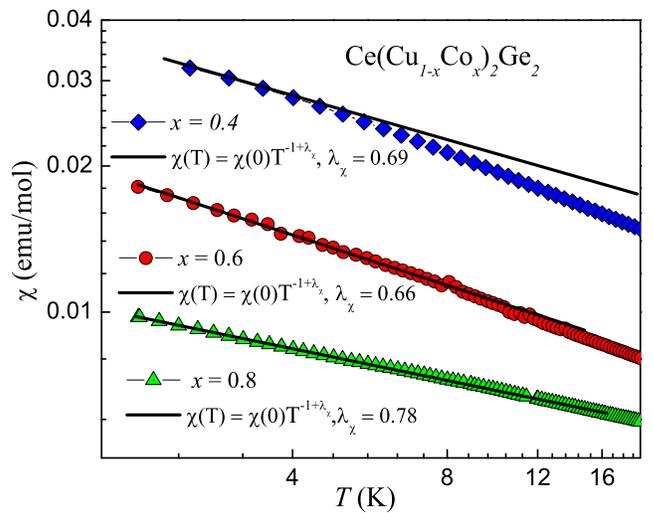}
\caption{\label{fig:CeCuCoGe-SUSP} (Color online) Temperature dependence of the magnetic susceptibility of  Ce(Cu$_{1-x}$Co$_x$)$_2$Ge$_2$ for $x$ = 0.4, 0.6, and 0.8.}

\end{figure}

\begin{table}
\centering
\caption [Crystallographic and Magnetic parameters of doped Ce(Cu$_{1-x}$Co$_x$)$_2$Ge$_2$ compounds]{Effective paramagnetic moments $\mu$$_{eff}$($\mu_{B}$), antiferromagnetic ordering temperature $T$$_N$(K), Curie-Weiss temperature $\theta$$_P$ and  Kondo temperature $T$$_{K}$(K) obtained from susceptibility ($T_{K}^{\chi}$(K)), magnetoresistance scaling ($T_{K}^{MR}$(K)), and entropy ($T_{K}^{S}$(K))of Ce(Cu$_{1-x}$Co$_x$)$_2$Ge$_2$}.
\label{table:CeCuCoGe-table3}
\vskip .3cm
\addtolength{\tabcolsep}{+2pt}
\begin{tabular}{c c c c c c c }
\hline
\hline
$x$ & $\mu_{eff}$($\mu_{B}$) & $\theta$$_P$(K)  & $T_{N}$(K) & $T_{K}^{\chi}$(K)& $T_{K}^{MR}$(K) & $T_{K}^{S}$(K) \\[0.5ex]
\hline
0 	&  2.50	& -25.2 & 4.1 & 6.3 & - & 7  \\[1ex]

0.02 &  2.54	& -26.2 & 3.0 & 6.5 & - & - \\[1ex]

0.05 &  2.66	&-30.2 & 2.1 &  7.5 &  6.7 &  6 \\[1ex]

0.1 &  2.59	&-35.8 &  - &  8.9 &  - &  6 \\[1ex]

0.15 &  2.71	&-37.7& - &  9.4 &  7.8 &  - \\[1ex]

0.2 &  2.59	&-35.8 &  0.6 &  8.9 &  8.5 &  8 \\[1ex]

0.40 &  2.8	    &-60.0   & - &  15.0 & - &  13 \\[1ex]

0.6 &  2.53	    &-105.4 & - &  26.4 &  24.7 &  19 \\[1ex]

0.8 &  2.61	    &-156.4 & - &  39.0 &  36.3 &  $\approx$ 24 \\[1ex]

1 & 2.69 &-399.8 & - &  99.7 &  - &  $>$ 50 \\ [1ex]
\hline
\end{tabular}
\end{table}

%\subsection*{\label{ExpDetails} C. Heat Capacity}

\begin{figure}
\includegraphics[width=8.5cm, keepaspectratio]{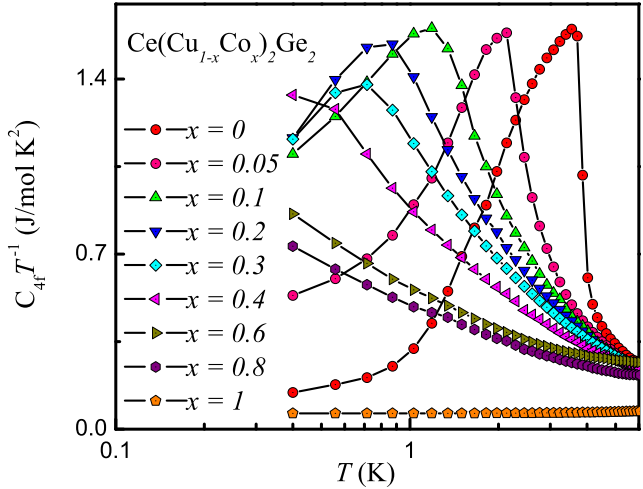}
\caption{\label{fig:CeCuCoGe-SPHT} (Color online) Low temperature heat capacity $C$$_{4f}$ divided by $T$ vs log$T$ of Ce(Cu$_{1-x}$Co$_x$)$_2$Ge$_2$ for $x$ = 0, 0.05, 0.1, 0.2, 0.3, 0.4, 0.6, 0.8, and 1 in the temperature range of 0.4 to 6 K.}
\end{figure}

\begin{figure}
\includegraphics[width=8.5cm, keepaspectratio]{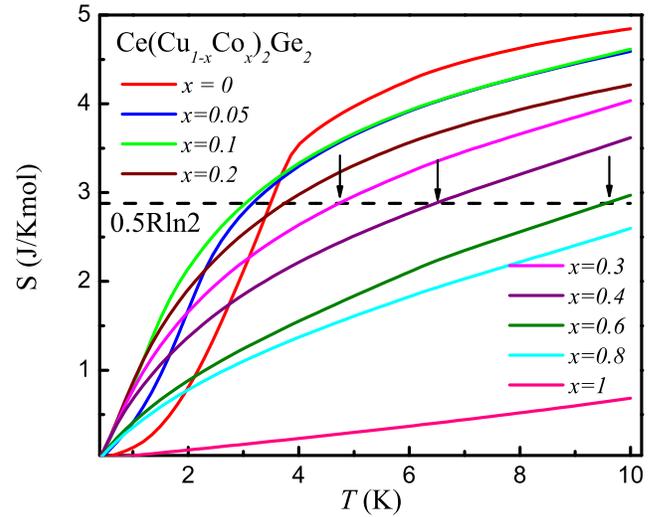}
\caption{\label{fig:CeCuCoGe-SPHT} (Color online) The 4f (magnetic) entropy($S$$_{mag}$) as a function of temperature  obtained from the heat capacity data as described in the text.}
\end{figure}

The magnetic part of the heat capacity $C$$_{4f}$($T$) was deduced by subtracting the heat capacity of LaCu$_2$Ge$_2$ and LaCo$_2$Ge$_2$ from that of Ce(Cu$_{1-x}$Co$_x$)$_2$Ge$_2$ after adjusting the renormalization to account for the slight atomic mass difference between La, Ce, Co and Cu, as follows:

\begin{equation}
\begin{split}
&C_{mag}[x]=C_P[x]-(1-x)\times C_P[LaCu_2Ge_2]
\\
&-(x)\times C_P[LaCo_2Ge_2]
\end{split}
\end{equation}

Figure 5 shows $C_{4f}/T$ vs $T$ for $x$ = 0, 0.05, 0.1, 0.2, 0.3 0.4, 0.6, 0.8, and 1. Low temperature anomaly in $C$$_{4f}$($T$) is associated with antiferromagnetic transition $T$$_{N}$ for 0 $\leq$ $x$ $\leq$ 0.4. For $x$ $>$ 0.4, the specific heat exhibits no anomaly down to 0.4 K. As $T$$_{N}$ approaches zero around $x$$_c$ = 0.6, $C_{4f}/T$ diverges down to 0.4 K, the lowest temperature at which the data were recorded. This is a common feature of non-Fermi-liquid behavior near a QCP in correlated $f$ -electron materials and associated with quantum critical fluctuation of the magnetic order parameter. The magnetic contribution to the entropy $S$$_{mag}$, calculated by integrating the $C$$_{mag}$/$T$ versus $T$, is shown in the Fig.6. The value of entropy for $x$ = 0.00 is 0.6 $R$ $\ln$2  at $T$ = 4 K and 0.8 $R$ $\ln$2 at 10 K. The reduced value of magnetic entropy suggests the presence of Kondo screening of the $f$ moment by the conduction electrons even in the magnetically ordered state \cite{Tran}. The full entropy expected for the $J$ = 5/2 multiplet of  Ce$^{3+}$ is recovered at room temperature \cite{Felten}. The black arrows indicate the position of Kondo temperature($T$$_K$/2) estimated using the relation $T$$_K$ = 2$\cdot$$T$($S$ = 0.5 $R$ ln$2$) \cite{Klingner} and the obtained values are listed in table I. The $C/T$ vs $T$ data for Ce(Cu$_{0.4}$Co$_{0.6}$)$_2$Ge$_2$, located at the magnetic-nonmagnetic boundary, is shown in the main panel of Fig.7. The upper right inset of Fig.7 shows the $C/T$ vs $T$ data for Ce(Cu$_{0.2}$Co$_{0.8}$)$_2$Ge$_2$ on log-log plot. The data for both $x$ = 0.6 and 0.8 have been fitted with power law $C/T$ = a$T$$^{-1+\lambda_C}$ in the temperature range 0.4 K $\leq$ $T$ $\leq$ 4 K and the obtained values of $\lambda_C$ are listed in table II. We note that there is a discrepancy in the values of $\lambda_C$ inferred from the fits to magnetization and heat capacity data. Similar discrepancies have also been observed by Castro Neto \cite{Andrade}, which were attributed to magneto crystalline anisotropy and preferred crystalline orientation in polycrystalline samples. In order to provide a direct comparison between power law and logarithmic behavior at critical concentration $x$$_c$, $C/T$ vs $T$ is presented in the lower left inset of Fig.7 on logarithmic scale. A logarithmic divergence corresponding to 2D fluctuations has also been observed experimentally for several NFL systems in the crossover regime near a AFM QCP \cite{Andrade,Scheidt}. From the lower inset of Fig.7 it is clear that the data follow the function $C/T$ = - a $ln(T)$ in comparatively small temperature range 0.4 K $\leq$ $T$ $\leq$ 1.0 K which is not entirely convincing. Our data for $x$ = 0.6 is also in marked contrast to the asymptotic ($T$$\rightarrow$0) dependencies predicted by the spin-fluctuation theory at the AFM QCP in 3D \cite{Moriya,Lohneysen}, namely, $C/T$ $\propto$ 1 - a$\sqrt{T}$. Thus, for concentrations near to $x$ $\sim$ 0.6, an  AFM QCP  is observed in this series and NFL behavior becomes evident as inferred from power law dependence over a significant temperature range. For CeCo$_2$Ge$_2$ we obtain $\lambda$ = 1, as expected for a Fermi liquid behavior.

\begin{figure}
\includegraphics[width=8.5cm, keepaspectratio]{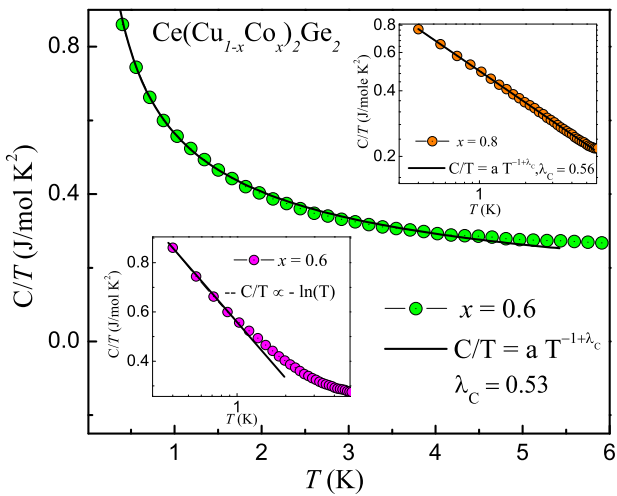}
\caption{\label{fig:CeCuCoGe-SPHT} (Color online) Low temperature specific heat of CeCu$_{0.8}$Co$_{1.2}$Ge$_2$. Upper right inset shows the specific heat of CeCu$_{0.2}$Co$_{1.6}$Ge$_2$ on log-log plot. The solid line are fit to the data with $C/T$ $\propto$ $T$$^{-1+\lambda}$ behavior. Lower inset shows $C$$_{4f}$/$T$ vs the logarithm of $T$ for CeCu$_{0.8}$Co$_{1.2}$Ge$_2$ and is fitted by $C$$_{4f}$/$T$ $\sim$ - $ln(T)$ (solid line).}
\end{figure}

\begin{figure}
\includegraphics[width=8.5cm, keepaspectratio]{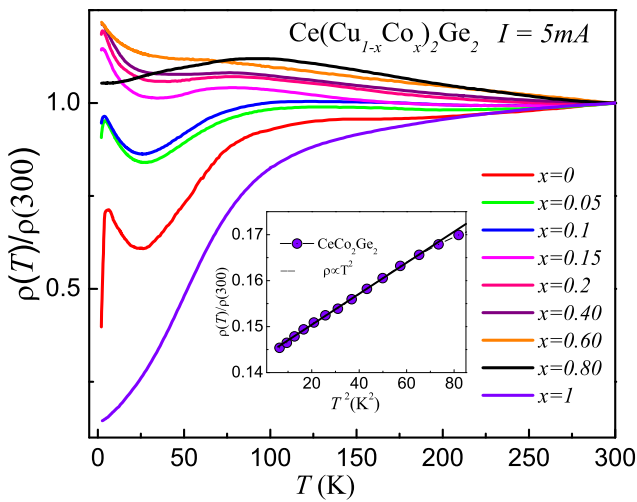}
\caption{\label{fig:CeCuCoGe-RT} (Color online) Temperature dependence of the resistivity normalized to the room-temperature value for 0 $\leq$ $x$ $\leq$ 1. Inset shows that the normalized resistivity of CeCo$_2$Ge$_2$ follows Fermi liquid behavior (Solid line) in the temperature range 2 K $\leq$ $T$ $\leq$ 8 K.}
\end{figure}

\begin{table}
\centering
\caption [Crystallographic and Magnetic parameters of doped Ce(Cu$_{1-x}$Co$_x$)$_2$Ge$_2$ compounds]{Exponent $\lambda$ obtained from fits with power laws $C/T$ = a$T$$^{-1+\lambda}$ to specific heat ($\lambda$$_C$) and magnetic susceptibility ($\lambda$$_\chi$) data of Ce(Cu$_{1-x}$Co$_x$)$_2$Ge$_2$ for $x$ = 0.4, 0.6, and 0.8}.
\label{table:CeCuCoGe-table2}
\vskip .8cm
\addtolength{\tabcolsep}{+15pt}
\begin{tabular}{c c c c }
\hline
\hline
$x$ & 0.4 & 0.6  & 0.8  \\[0.5ex]
\hline
$\lambda$$_C$ 	&  -	& 0.53 & 0.56  \\[1ex]

$\lambda$$_\chi$ &  0.69	& 0.66 & 0.78 \\[1ex]

\hline
\end{tabular}
\end{table}

%\subsection*{\label{ExpDetails} D. Electrical Resistivity}

Figure 8 shows the temperature dependence of normalized electrical resistivity of Ce(Cu$_{1-x}$Co$_x$)$_2$Ge$_2$ in the range 0 $\leq$ $x$ $\leq$ 1. The broad but well-defined maxima at around $T$$_{CF}$ = 100 K is due to the crystal field (CF) effect. The low-temperature maxima ($T$$_{max}$) at around 6 K  can be attributed to Kondo coherence. It is monotonically decreasing with increasing $x$ in sharp contrast to increase in $T$$_{max}$ in CeCu$_2$(Si$_{1-x}$Ge$_x$)$_2$ \cite{Knebel} and CeCu$_2$Ge$_2$ under pressure \cite{Honda}. For the Co doped samples the resistivity at 2 K is approximately the same as that around 300 K and we did not observe large resistance drop associated with Kondo coherence atleast down to 2 K. We believe that the decrease in the value of RRR is due to dominating Kondo type scattering at low temperature as in the case of CePd$_{1-x}$Rh$_x$ \cite{Sereni} and Ce(Pd$_{1-x}$Ni$_x$)$_2$P$_2$ \cite{Lai}. Low temperature resistivity data of reference \cite{Maeda} confirms the deviation from FL for $x$ = 0.6 whereas Ce(Cu$_{1-x}$Co$_x$)$_2$Ge$_2$ recovers its FL nature for $x$ $\geq$ 0.8 \cite{Maeda,Fujii} where the resistivity follows a quadratic temperature dependence $\rho$($T$) - $\rho$($0$) = $\Delta$$\rho$ = $A$$T$$^2$ (inset of Fig.8).

One can estimate Kondo temperature by carefully analyzing magnetoresistance (MR) data. It is clear from previous studies on CeCu$_2$Ge$_2$ \cite{Zeng} that the magnetoresistance is positive in the magnetically ordered state, whereas it is negative in the paramagnetic state. The positive magnetoresistance in the ordered state is consistent with the antiferromagnetic nature of the magnetic ordering. In the paramagnetic region, the negative magnetoresistance is due to the freezing out of spin-flip scattering in a Kondo compound by the magnetic field. Figure 9 represents  normalized magnetoresistance  measured in the paramagnetic state  plotted as a function of  $\mu$$_0$H/($T$+$T$*) for $x$ = 0.05, 0.15, and 0.2 which allows us to map MR data measured at different temperatures (well above AFM ordering ) onto a single curve. Here, $T$* is the characteristic temperature which is an approximate measure of the Kondo temperature($T$$_K$) \cite{Moodenbaugh, Hossain}. Thus estimated values of $T$$_K$ for different concentrations are in good agreement with the $T$$_K$ values inferred from magnetic susceptibility and heat capacity data and they are listed in table I.

\begin{figure}
\includegraphics[width=8.5cm, keepaspectratio]{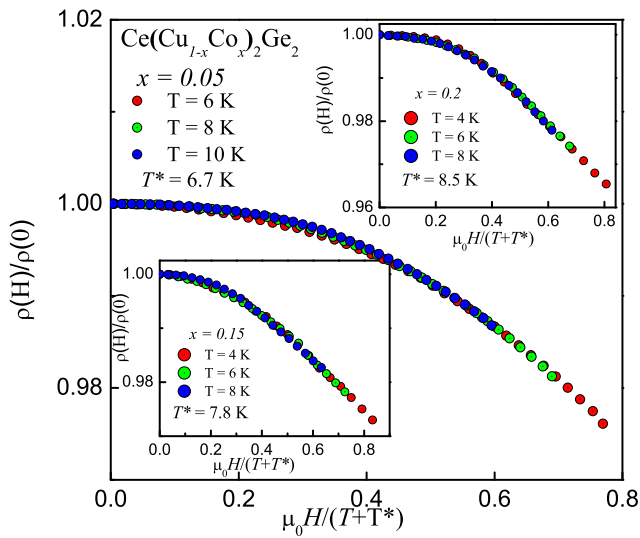}
\caption{\label{fig:CeCuCoGe-RH} (Color online) Normalized resistivity plotted as a function of $\mu$$_0$H/($T$+$T$*) for $x$ = 0.05, 0.15, 0.2, where $T$* is the characteristic temperature.}
\end{figure}

\begin{figure}
\includegraphics[width=8.5cm, keepaspectratio]{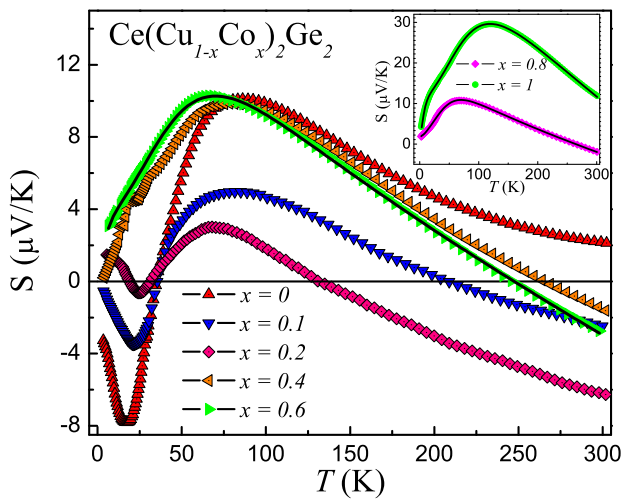}
\caption{\label{fig:CeCuCoGe-THP} (Color online) Temperature dependence of thermoelectric power for $x$ = 0, 0.1, 0.2, 0.4 and 0.6. The thermopower of $x$ = 0.8 and 1 is shown in inset. Solid lines are fits to the data using Eq.3.}
\end{figure}

%\subsection*{\label{ExpDetails} E. Thermopower}

The temperature-dependent thermopower $S$($T$) of Ce(Cu$_{1-x}$Co$_x$)$_2$Ge$_2$ for $x$ = 0, 0.1, 0.2, 0.4, 0.6 is shown in Fig.10. The data for CeCu$_2$Ge$_2 $ is in good agreement with the literature \cite{Behnia}. $S$($T$) for $x$ = 0, 0.1 and  0.2 shows a broad maxima around 90 K along with a sign change at 34 K and a minima with the negative value of Seebeck coefficient equal to - 8 $\mu$V/K for $x$ = 0.00 and - 2.5 $\mu$V/K for $x$ = 0.10. The negative peak in the thermopower below 30 K is attributed to Kondo scattering on the crystal-field ground state \cite{Behnia}. It becomes less pronounced with increasing $x$ and for $x$ = 0.4, 0.6, and 0.8 (inset of Fig.10), we observed only the broad maxima. The thermopower is positive and significantly enhanced for $x$ = 1 (inset of Fig.10), which is found in several Ce-based intermediate valance systems like CeNi$_2$Si$_2$ \cite{Sampa,Levin} and CePd$_3$ \cite{Besnus}. A similar feature in thermopower is also seen for CeCu$_2$Ge$_2$ under pressure \cite{Link}, where low temperature negative peak disappears and becomes positive in the pressure range of 7.8 GPa to 11.2 GPa. It is important to note that in the $p$-$T$ phase diagram of CeCu$_2$Ge$_2$ the disappearance of AFM order and emergence of superconducting  phase has been found in the same pressure range \cite{DJaccard}. Furthermore, thermoelectric properties of many Ce and Yb - based intermediate valance system is well described using a phenomenological valence-fluctuation model \cite{Wis,Garde,Fre}. In this model, a Lorentzian shaped $4f$ band is located at the energy $\epsilon$$_f$  ($k$$_B$$T$$_0$) below the Fermi level, where $T$$_0$ is temperature independent parameter in the intermediate valance regime.  Width of the band $\Gamma$, which is proportional to the number of states that would effectively take part in the scattering process, depends on temperature as $\Gamma$ = $T$$_f$ exp(-$T$$_f$/$T$). Here $T$$_f$ is a parameter related to the quasielastic linewidth, arising from the hybridization between the $4f$ electrons (forming a narrow band) and the surrounding conduction electrons (forming a broad band). The thermopower can be described by the function:

\begin{equation}
S(T)=\frac{C_1 T_0 T}{T_0^2+\Gamma (T)^2} + C_2 T
\end{equation}

Where C$_1$ and C$_2$ are temperature-independent parameters, which determine the strength of the contributions from the non-magnetic and magnetic scattering processes, respectively. Now, $S(T)$ data for $x$ = 1, 0.8, and  0.6 can't be modeled by the Eq.2 due to the presence of an additional hump like feature below 50 K.  Therefore we used an additional quasiparticle-like term \cite{Got} given by the formula $S(T)$ = $AT$/($B$ + $T$$^2$), where A = 2$\epsilon$$_f$/$|$e$|$ and B = 3($\epsilon$$_f$$^2$ + $\Gamma$$^2$)/($\pi$$^2$$k$$_B$$^2$) are the temperature independent parameters. Therefore, the total $S(T)$ could then be expressed as

\begin{equation}
S(T)=\frac{C_1 T_0 T}{T_0^2+\Gamma (T)^2} + C_2 T + \frac{A T}{B + T^2}
\end{equation}

Eq. 3 well replicates the observed $S(T)$ data  for $x$ = 1, 0.8 (inset of Fig.10), and  0.6 (Fig.10). The parameter $T$$_f$ increases to 95 K and 103 K for $x$ = 0.6 and 0.8 respectively and afterwards to even larger value of 164 K for $x$ =1. Furthermore, the value of $T$$_0$ is 95 K for $x$ = 1 where as for $x$ = 0.8 and 0.6 it has nearly the same value of 47 K. These results suggest that the cerium valence evolves away from a purely trivalent state which is consistent with the deviation from Vegard's law of lattice volume and hump like feature of inverse susceptibility for $x$ $\geq$ 0.6. More detailed study using XANES measurements is needed to determine the valance evolution of Ce with doping level.

\begin{figure}
\includegraphics[width=8.5cm, keepaspectratio]{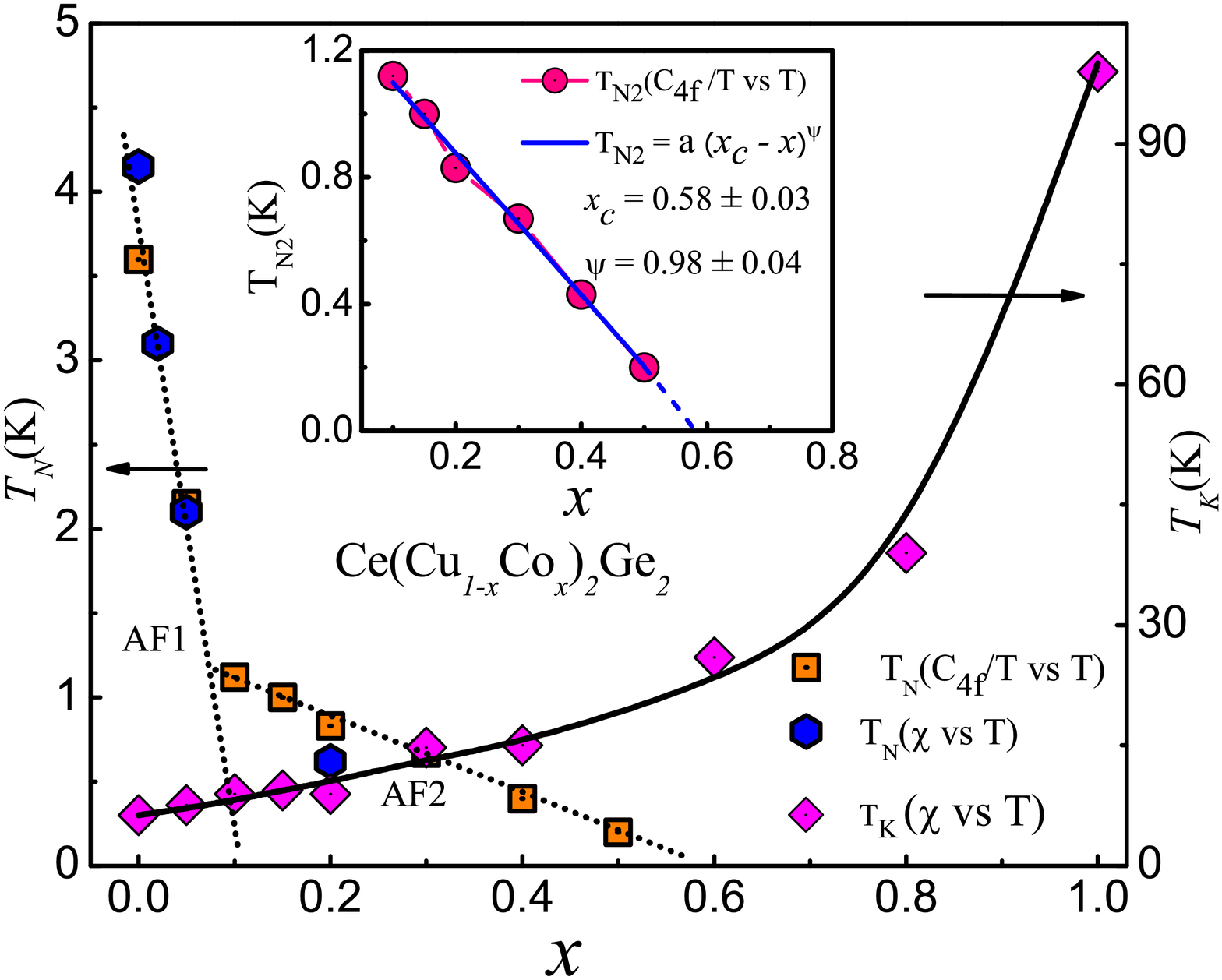}
\caption{\label{fig:CeCuCoGe-PD} (Color online) Variation of AFM ordering temperature $T$$_N$ (left panel) and Kondo temperature $T$$_K$ (right panel) as a function of Co doping $x$ . Inset shows $T$$_{N2}$ vs $x$ data fitted with $T$$_{N2}$ = a ($x$$_c$ - $x$)$^\psi$ (solid line) in the range 0.1 $\leq$ $x$ $\leq$ 0.5, expected for 2D quantum critical fluctuations.}
\end{figure}

Our results of electrical transport, magnetic susceptibility, heat capacity and thermopower measurements lead to a consistent picture of the magnetic behavior of the polycrystalline Ce(Cu$_{1-x}$Co$_x$)$_2$Ge$_2$. The $x$-$T$ phase diagram is presented in Fig.11, where $T$$_N$ shows two different slopes for 0 $\leq$ $x$ $<$ 0.1 (AF1, $T$$_{N1}$) and 0.1 $\leq$ $x$ $\leq$ 0.6 (AF2, $T$$_{N2}$). In the phase diagram, the point corresponding to $C$/$T$ for $x$ = 0.5 is taken from the reference \cite{Maeda}. Pure CeCu$_2$Ge$_2$ also reveals two different magnetically ordered phase under external pressure\cite{Honda}. In order to determine the effect of pressure (chemical) on $T_N(x)$ dependence, we can compare the lattice parameters of CeCu$_2$Ge$_2$ under chemical pressure (i.e. of Ce(Cu$_{1-x}$Co$_x$)$_2$Ge$_2$) with those of CeCu$_2$Ge$_2$ under hydrostatic pressure. We found that the volume of $x$ = 0.6 sample, where $T_N$ goes to zero, is equal to that of CeCu$_2$Ge$_2$ at 2.5 GPa. However, the hydrostatic pressure vs $T_N$ phase diagram does not show any appreciable change in $T_N$ up to the pressure of 2.5 GPa. This indicates that in Ce(Cu$_{1-x}$Co$_x$)$_2$Ge$_2$ the carrier concentration modification play the major role in the suppression of magnetic ordering. The phase diagram of Ce(Cu$_{1-x}$Ni$_x$)$_2$Ge$_2$ \cite{Loidl} also shows two distinct types of antiferromagnetic ordering, representing heavy-fermion band magnetism (HFBM) and local-moment magnetism (LMM). So, in our case one can presume that the different slopes of $T$$_N$ vs $x$ in different concentration range are due to the different kind of magnetic ordering (local and itinerant), which requires further confirmation. Furthermore, in the $x$-$T$ phase diagram near a QCP the N$\acute{e}$el temperature varies as $T$$_N$ $\sim$ $\mid$$x$$_c$ - $x$$\mid$$^\psi$ with $\psi$ = $z/(d+z-2)$, where $x$ is the doping concentration and $z$, a dynamic critical exponent relating the length and time scales of critical fluctuations \cite{Castro,Millis,Mathur}. The value of $z$ is expected to be $2$ and $3$ for AFM and ferromagnetic (FM) QCP respectively. The value of $d$ equals $3$ and $2$  for 3D and 2D critical fluctuations respectively. In the inset of Fig.11, the solid line shows $T$$_{N2}$ = a ($x$$_c$ - $x$)$^\psi$  with  $x$$_c$ $\sim$ 0.58 $\pm$ 0.03 and $\psi$ = 0.98 $\pm$ 0.04 by fitting with the data of $T$$_{N2}$ vs $x$ for  Ce(Cu$_{1-x}$Co$_x$)$_2$Ge$_2$ (0.1 $\leq$ $x$ $\leq$ 0.5). The linear behavior of $T$$_N$ with ($x$$_c$ - $x$) is consistent with 2-dimensional nature of quantum critical fluctuation in this system. Another important observation near QCP is the NFL behavior. In order to discuss this behavior, we have to take into account that two effects occur simultaneously in our system. One concerns the hole doping on Cu site, which tunes the relative strengths of the Kondo and RKKY interactions, and the other manifests disorder effect through alloying. We anticipate that the combined behavior, i.e. the competition between the Kondo effect and the RKKY interaction in presence of disorder, could result in the formation of magnetic clusters in the proximity to the QCP leading to NFL behavior which is consistent with the predictions of the model proposed by Castro Neto et al. The analysis of the $C$$_{4f}$/$T$ and $\chi$($T$) suggests NFL behavior, where a power-law dependence of $C/T$ = a$T$$^{-1+\lambda}$ have been found for $x$ = 0.6, and 0.8. The non-Fermi-liquid effects in the specific heat and  dc susceptibility is compatible with the quantum Griffiths phase scenario.

\section{Conclusion}

We have reported a comprehensive study of electrical transport, magnetic susceptibility, heat capacity and thermopower measurements on Co doped CeCu$_2$Ge$_2$. The $T$$_{N}$ vs $x$ phase diagram reveals two distinct regimes that might be related to two different kinds of magnetic order. A significant deviation of physical properties from a FL behavior such as $\Delta$ $\rho$ $\propto$ $T$$^{2\pm\delta}$ and  $C/T$ $\propto$ $\chi$($T$) $\propto$ $T$$^{-1+\lambda}$ are observed around $x$$_c$ = 0.6 and attributed to an AF-QCP with $T$$_N$ = 0. The 2D nature of quantum fluctuations is inferred from magnetic phase diagram where $T$$_{N2}$ follows $T$$_{N2}$ $\sim$ ($x$$_c$ - $x$) behavior. We have been able to disentangle the relative
importance of the influence of volume change, carrier concentration change and disorder (Kondo disorder arising out of small variation of the Cu/Co concentration) on the physical properties. We find that the rapid decrease of $T_N$ upon Co-doping is mainly due to carrier concentration change and associated change of the $T_K$ and $T_{RKKY}$. The disorder plays an important role in deciding the nature of the phase around magantic-nonmagnetic boundary. Instead of standard Quantum Critical spin density wave (SDW) found in pure heavy Fermion antiferromagnetic compounds, we found that Griffiths phase is stabilized around the critical concentration. To get more insight, experiments on single crystal are desirable. Neutron diffraction measurements are required to confirm the exact nature of magnetic ordering of the doped compounds.

\subsection*{\label{ExpDetails} ACKNOWLEDGEMENTS}

We gratefully acknowledge Christoph Klausnitzer for his assistance during low temperature magnetization measurements and J.D. Thompson for stimulating discussion.

\end{document}